\documentclass[a4paper]{jpconf}
\usepackage{graphicx}
\begin{document}
\title{Heavy quark energy loss and thermalization in hot and dense nuclear matter}

\author{Shanshan Cao, Guang-You Qin and Steffen A. Bass}

\address{Department of Physics, Duke University, Durham, North Carolina 27708, USA}

\ead{shanshan.cao@duke.edu}

\begin{abstract}
We study  heavy quark energy loss and thermalization in hot and dense nuclear medium. The diffusion of heavy quarks is calculated via a Langevin equation, both for a static medium as well as for a quark-gluon plasma (QGP) medium generated by a (3+1)-dimensional hydrodynamic model. We investigate how the initial configuration of the QGP and its properties affect the final state spectra and elliptic flow of heavy flavor mesons and non-photonic electrons. It is observed that both the geometric anisotropy of the initial profile and the flow profile of the hydrodynamic medium play important roles in the heavy quark energy loss process and the development of elliptic flow. Within our definition of the thermalization criterion and for reasonable values of the diffusion constant, we observe thermalization times that are longer than the lifetime of the QGP phase.
\end{abstract}

\section{Introduction}

It is now generally accepted that a deconfined state of QCD matter, the {\it strongly interacting quark-gluon plasma} (sQGP) is created during relativistic heavy-ion collisions at RHIC and LHC \cite{Gyulassy:2004zy,Harris:1996zx}. This highly excited state of matter displays properties similar to a nearly perfect fluid, which has been successfully described by hydrodynamics models \cite{Gyulassy:2004zy,Nonaka:2006yn,Song:2010mg}.

Heavy quarks, being majorly produced in the very early stage of heavy-ion collisions and then transported through the dense medium, serve as hard probe of QGP properties. In the past decade, experimental observations have revealed a great many interesting and sometimes surprising results for heavy flavors, such as the large value of elliptic flow $v_2$ and small value of nuclear modification factor $R_{AA}$ of heavy mesons \cite{Bianchin:2011fa,Rossi:2011nx} and non-photonic electrons \cite{Adare:2010de}, indicating a stronger coupling between heavy quarks and the QGP medium than expected. Therefore, it would be of great interest to carry out a systematic study of the heavy quark energy loss process inside such a hot and dense nuclear matter and verify whether they indeed approach equilibrium during the QGP lifetime due to such strong coupling.

Between the two energy loss mechanisms, gluon radiation can be greatly suppressed by the large mass of heavy quark because of the ``dead-cone effect" \cite{Dokshitzer:2001zm}, and therefore, quasi-elastic scattering off light quarks and gluons \cite{Mustafa:2004dr} is usually considered as the dominant factor for heavy quark evolution at RHIC energies. In the limit of multiple interactions where the energy transfer during each interaction is small, the heavy quark propagation through a QGP matter can be treated as the Brownian motion and therefore be described by the Langevin equation \cite{Svetitsky:1987gq,Moore:2004tg,He:2011qa,Young:2011ug}. We shall apply such Langevin framework to heavy quark evolution inside both a static medium with fixed temperature and a realistic QGP matter modeled with a (3 + 1)-dimensional relativistic hydrodynamic approach \cite{Nonaka:2006yn}, and investigate how heavy flavor spectra observed in relativistic heavy-ion collisions depend on various ingredients in the phenomenological studies, such as the initial production of heavy quarks, the geometry and the flow properties of the hydrodynamic medium, and the coupling strength between heavy quarks and medium. Furthermore, we attempt to define a rigorous criterion of heavy quark thermalization and answer the question of whether general features seen in the data, such as the presence of elliptic flow and a small value of the nuclear modification factor can be used to conclude that heavy quarks actually have thermalized in the QGP, or they remain off-equilibrium during their entire evolution, despite exhibiting a strong response to the surrounding medium.

This paper is based on our previous work \cite{Cao:2011et,Cao:2012jt} and will be organized as follows. In the next section, we will discuss the methodology used to study heavy quark production, evolution in medium and hadronization, and introduce our criterion for testing the thermalization of heavy quarks. In Sec.\ref{thermalizationprocess}, we will investigate the thermalization process of charm quarks in a static medium as well as in a cooling and expanding QGP medium. In Sec.\ref{modeldep},  we will systematically explore how experimental observables, such as heavy flavor suppression and flow, depend on various phenomenological inputs. A summary and outlook will be provided in the last section.

\section{Methodology}
\label{methodology}

\subsection{Heavy quark evolution}

In this work, we shall only consider the collisional energy loss experienced by heavy quarks while traveling through the QGP. The radiative energy loss will be incorporated in a follow-up study. In the limit of multiple quasi-elastic scatterings, heavy quark diffusion inside a thermalized medium can be treated in the framework of Langevin equation:
\begin{equation}
\frac{d\vec{p}}{dt}=-\eta_D(p)\vec{p}+\vec{\xi}.
\end{equation}
In general, the noise term $\vec{\xi}$ may depend on the momentum of the particle, but for simplicity we assume that this is not the case here. The random momentum kicks satisfy the following correlation relation:
\begin{equation}
\langle\xi^i(t)\xi^j(t')\rangle=\kappa\delta^{ij}\delta(t-t').
\end{equation}
If the energy transfer during each interaction is small, the fluctuation-dissipation relation applies:
\begin{equation}
\eta_D(p)=\frac{\kappa}{2TE}.
\end{equation}
The diffusion coefficient is related to the drag term via:
\begin{equation}
D=\frac{T}{M\eta_D(0)}=\frac{2T^2}{\kappa}.
\end{equation}

We shall use the above Langevin equation to simulate heavy quark evolution inside a static as well as a dynamical QGP medium. For the former, the only required information on the medium is temperature, which remains fixed throughout the whole transport of heavy quarks. For an expanding QGP, we utilize a fully (3 + 1)-dimensional relativistic ideal hydrodynamic model \cite{Nonaka:2006yn}. The initial conditions of the hydrodynamic calculation are tuned to describe the hadronic data in the soft sector, such as hadron yields, spectra, and rapidity-distributions as well as radial and elliptic flow. Both Glauber and KLN-CGC models will be applied to our hydrodynamic simulation, and the difference between these two initializations will be compared in details later in terms of the influence on the final heavy flavor spectra. The hydrodynamic model provides us with the time evolution of the spatial distribution of temperature and collective flow velocity. Using our knowledge of the local flow velocity, for every Langevin time step we boost the heavy quark to the local rest frame of the fluid cell through which it propagates. In the local rest frame of the cell we perform the Langevin evolution using the local temperature before boosting back to the global c.m. frame. The evolution of heavy quark both in the pre-equilibrium phase before the formation of QGP (set to be 0.6~fm/c) and after the medium freeze-out are treated as free-streaming\footnote{For calculating experimental observables ($R_{AA}$ and $v_2$) in Sec.\ref{modeldep}, we treat the heavy quark motion below $T_c$ (around 160~MeV) as free-streaming before they fragment to mesons. However, to examine the time-evolution of heavy quark temperature parameter in a wider regime in Sec.\ref{thermalizationprocess}, we extend the Langevin evolution of heavy quark till the kinetic freeze-out (around 110~MeV).}.

Before their evolution, heavy quarks are initialized using Monte-Carlo Glauber model for the spatial distribution and the leading-order pQCD calculation for their momentum distribution. After they traverse the medium, the fragmentation of heavy quarks into heavy flavor mesons\footnote{The contribution of the coalescence mechanism to the hadronization process  has been shown to be significant in \cite{He:2011qa}, especially in the low $p_T$ region; this will be incorporated into our framework in a follow-up study. } and their decay into electrons are simulated via Pythia 6.4 \cite{Sjostrand:2006za}. Since there still exist uncertainties in the relative normalization of charm and bottom quarks \cite{Armesto:2005mz}, we shall treat this as a free parameter in simulation rather than fix it. And we will investigate the effect of this normalization on the quenching and the elliptic flow of heavy flavor decay electrons.

We select the final state particles in the mid-rapidity region ($-1<\eta<1$) and calculate the momentum space elliptic flow coefficient $v_2$ and the nuclear modification factor $R_{AA}$ as follows:
\begin{eqnarray}
&& v_2(p_T)=\langle \cos(2\phi)\rangle=\left\langle\frac{p_x^2-p_y^2}{p_x^2+p_y^2}\right\rangle, \nonumber\\
&& R_{AA}=\frac{\left({dN}/{dp_T}\right)_{\textnormal{fin}}}{\left({dN}/{dp_T}\right)_{\textnormal{init}}}.
\end{eqnarray}
Note that when heavy quarks are directly analyzed, the denominator and the numerator of $R_{AA}$ are the initial heavy quark distribution and the distribution of those surviving from the energy loss and passing through the medium. When analyzing heavy flavor mesons or electrons, the denominator represents the spectra of the corresponding particles fragmented/decayed directly from the initial heavy quarks, while the numerator represents those produced from the heavy quarks after transporting through the QGP medium. 

\subsection{Thermalization criterion}

Both the energy and momentum spectra will be utilized to study the thermalization process of heavy quarks. For a medium at fixed temperature without any inherent collective flow, an ensemble of thermalized
heavy quarks has the following energy distribution that allows for a straightforward extraction of its ``temperature'' via an exponential fit:
\begin{equation}
\label{efit}
\frac{dN}{pEdE}=C e^{-E/T}.
\end{equation}
While this particular form for the energy distribution provides a convenient representation for the extraction of the ``temperature'' of the heavy quark ensemble, it is insufficient to actually indicate thermalization, since we still need to verify the isotropy of particle momenta:
\begin{equation}
\label{}
f(p_i)\,=\, C \cdot T(\sqrt{p_i^2+m^2}+T)e^{-\sqrt{p_i^2+m^2}/T}.
\end{equation}
Note that if e.g., we initialize our heavy quark ensemble with a finite momentum along a given coordinate axis, its momentum distribution along that axis will be blue-shifted -- this can be taken into account by shifting the distribution along that axis using a parameter \begin{math} \tilde{p_i} \end{math}. For example, for the $z$ axis this
would give\footnote{Rigorously, $p_z$ should be boosted via $\gamma p_z+\gamma \beta E$. However, it is found that in our study, for $\beta$ not too large (below 0.8), the much more convenient Eq.(\ref{pfit}) already fits the spectrum well (with an error less than 5\% for $T$).}:
\begin{equation}
\label{pfit}
f(p_z)=C \cdot T(\sqrt{(p_z-\tilde{p_z})^2+m^2}+T)e^{-\sqrt{(p_z-\tilde{p_z})^2+m^2}/T}.
\end{equation}

For a realistic QGP medium, establishing thermalization requires the following procedure: for a given time step, we select all cells in our hydrodynamic medium within a temperature band of $T \pm \Delta T$. Then, we boost
all the heavy quarks localized in those cells into the respective local rest frames of the cells they are residing in and extract their temperature parameters from both their energy and momentum distributions. If these temperature parameters extracted from different ways agree with each other and meanwhile lie within the temperature band selected at the very beginning, we can conclude that the selected heavy quark ensemble has thermalized in the medium at the given temperature and the selected time step. This procedure can then be repeated for other temperature bands and time steps.

\section{Heavy Quark Thermalization}
\label{thermalizationprocess}

We start our investigation by simulating the diffusion of charm quarks in an infinite medium with a static temperature of 300~MeV. The charm quarks are initialized with 5~GeV/c momentum in the $\hat{z}$ direction, and the spatial diffusion coefficient is set to be $D = 6/(2\pi T)$, which is the value that was found in
\cite{Moore:2004tg} to be consistent with experimental observations \cite{Adare:2010de} in terms of non-photonic electron suppression and flow.

\begin{figure}[htb]
\begin{minipage}{18pc}
\includegraphics[width=18pc,clip=]{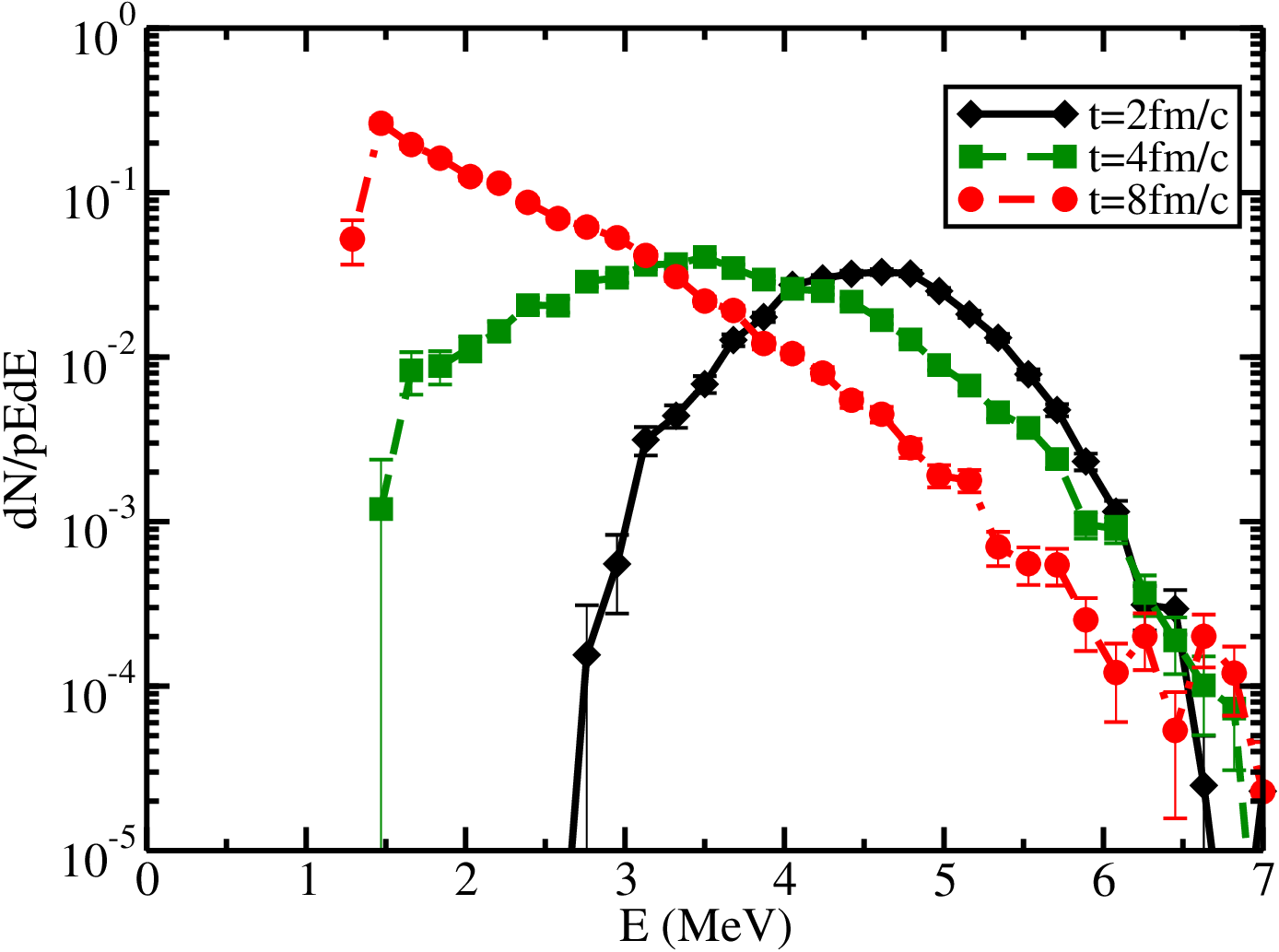}
\end{minipage}\hspace{2pc}
\begin{minipage}{14pc}
\includegraphics[width=18pc,clip=]{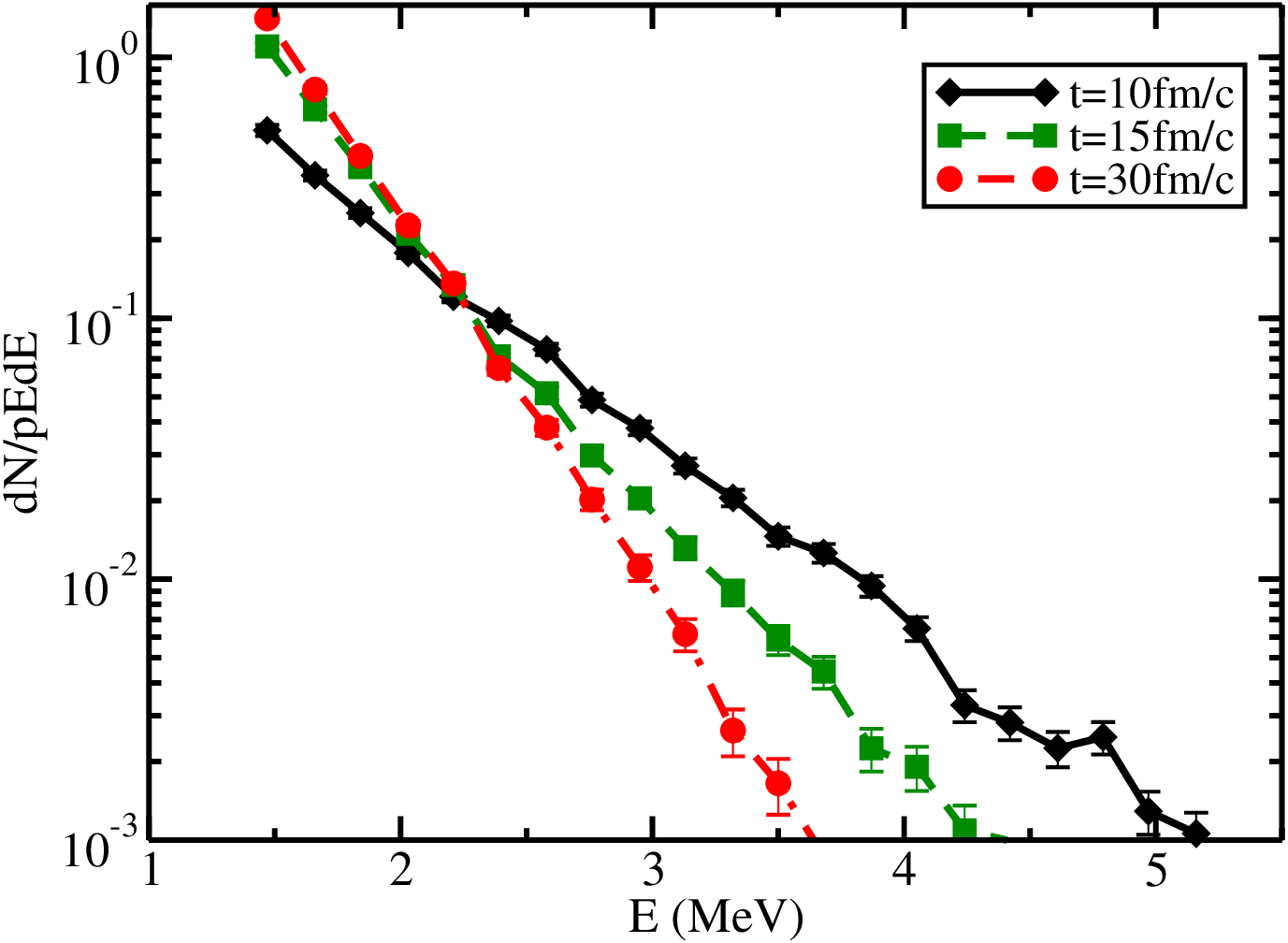}
\end{minipage}
\caption{\label{espectrum}(Color online) The evolution of the energy spectrum with respect to time. The left shows the results between 2~fm/c and 8~fm/c, where no linear relation is observed; and the right shows the results between 10~fm/c and 30~fm/c, where the linear relation is apparent.}
\end{figure}

Fig.\ref{espectrum} shows the energy spectrum $dN/pEdE$ vs. $E$ for different diffusion times. Before 10~fm/c, no linear relation can be observed between $\ln(dN/pEdE)$ and $E$. Such a linear relation occurs and the distribution appears thermal for longer time. However, the slope keeps increasing with respect to time and does not converge to the temperature of the medium until a diffusion time of around 30~fm/c. It indicates that the shape of the energy distribution alone is insufficient to conclude the thermalization of our ensemble of charm quarks.

\begin{figure}[htb]
\includegraphics[width=18pc,clip=]{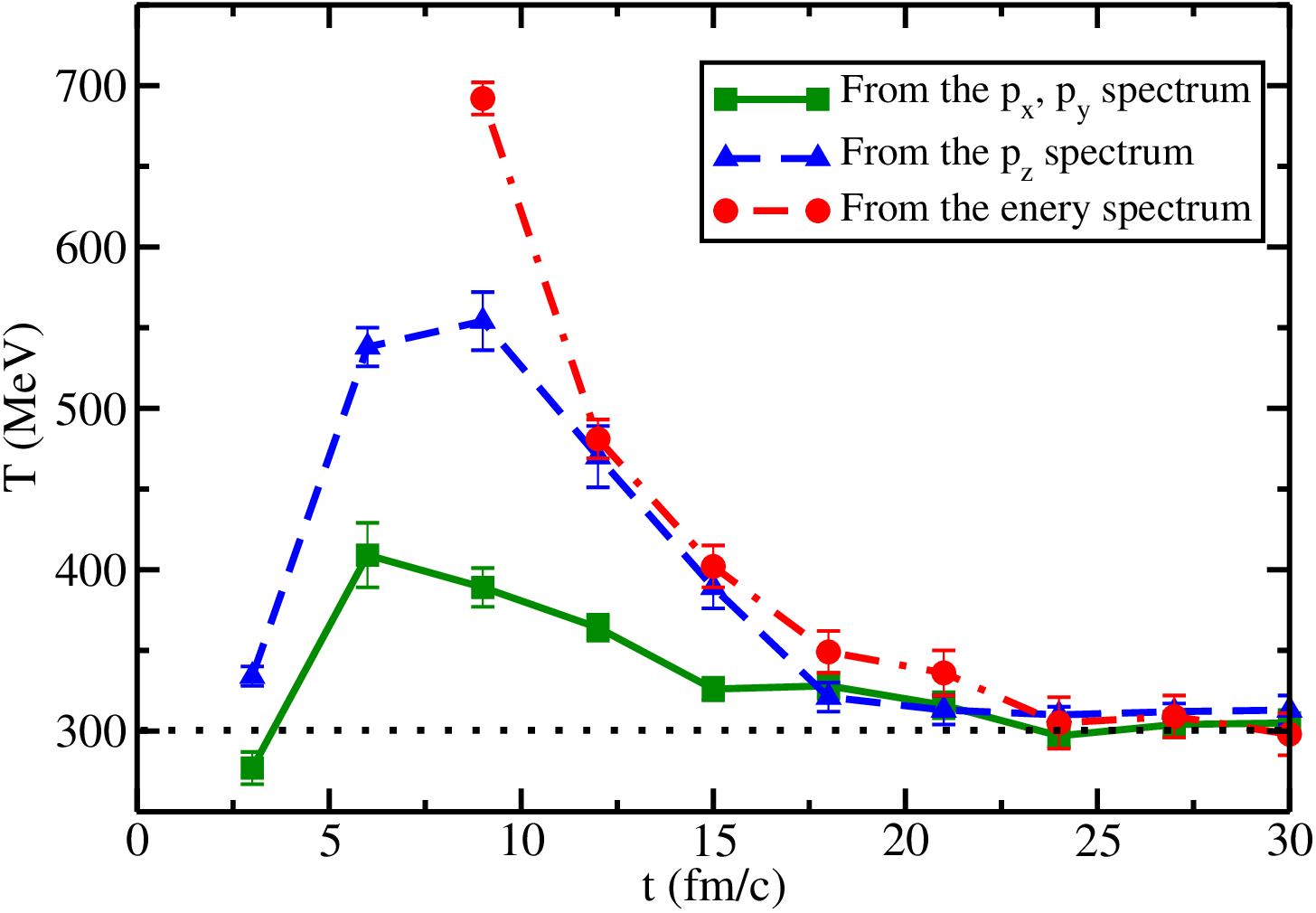}\hspace{2pc}
\begin{minipage}[b]{18pc}\caption{\label{Tcompare}(Color online) A comparison of ``temperatures'' obtained from $p_x$, $p_y$, $p_z$ and the energy spectra. Since $\hat{x}$ and $\hat{y}$ directions are symmetric, we take the mean value of the ``temperatures'' fitted from $p_x$ and $p_y$ spectra here.}
\end{minipage}
\end{figure}

We can directly extract the temperature parameters for charm quarks from these energy spectra and compare them with the temperature parameters extracted from momentum spectra, as shown in Fig.\ref{Tcompare}. Because the charm quarks are initialized in the $\hat{z}$ direction, their motion in the $x$-$y$ plane is majorly contributed by the thermal fluctuations and therefore the temperature parameters extracted from $p_x$ and $p_y$ distributions do not deviate significantly from the medium temperature. To the contrary, the charm quark motion in the $\hat{z}$ direction is blue-shifted because of our particular initialization, and consequently, the extracted temperature parameter from the $p_z$ spectrum does not coincide with that of the medium. Instead, it appears close to that from the energy spectrum and cools down together towards the medium temperature. We define the equilibrium as the point where the temperature parameters extracted from our different methods merge and approach that of the medium. In this scenario, it happens around the time of 25~fm/c.

With the above diffusion methodology and equilibrium criterion, we are able to study the thermalization process of heavy quarks in a realistic QGP medium simulated by a hydrodynamic calculation. The choice of parameters here are consistent with the 200~GeV Au-Au collisions at RHIC, and the impact parameter is set to be 2.4~fm.

\begin{figure}[htb]
\begin{minipage}{18pc}
\includegraphics[width=18pc,clip=]{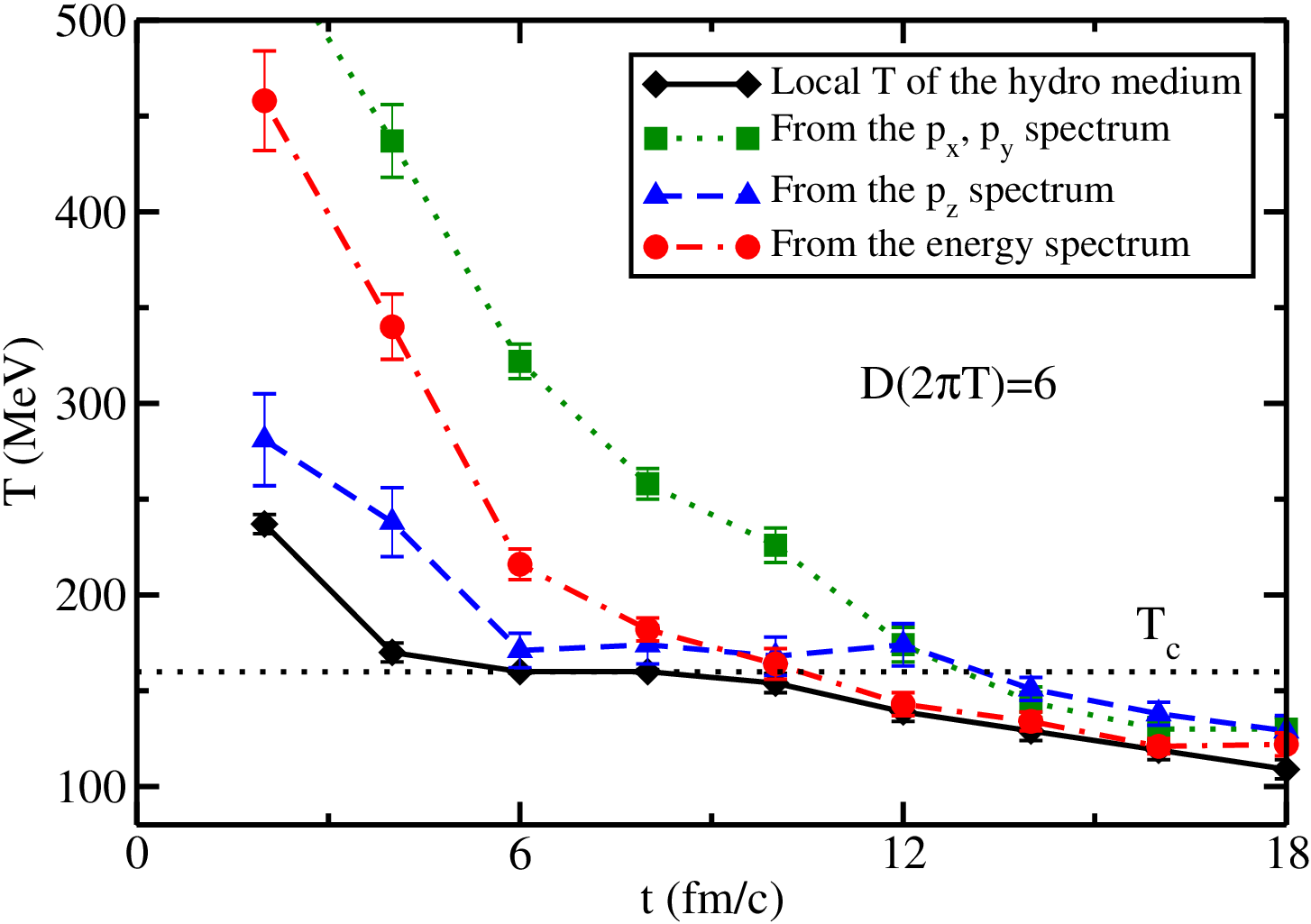}
\label{cascade6}
\end{minipage}\hspace{2pc}
\begin{minipage}{14pc}
\includegraphics[width=18pc,clip=]{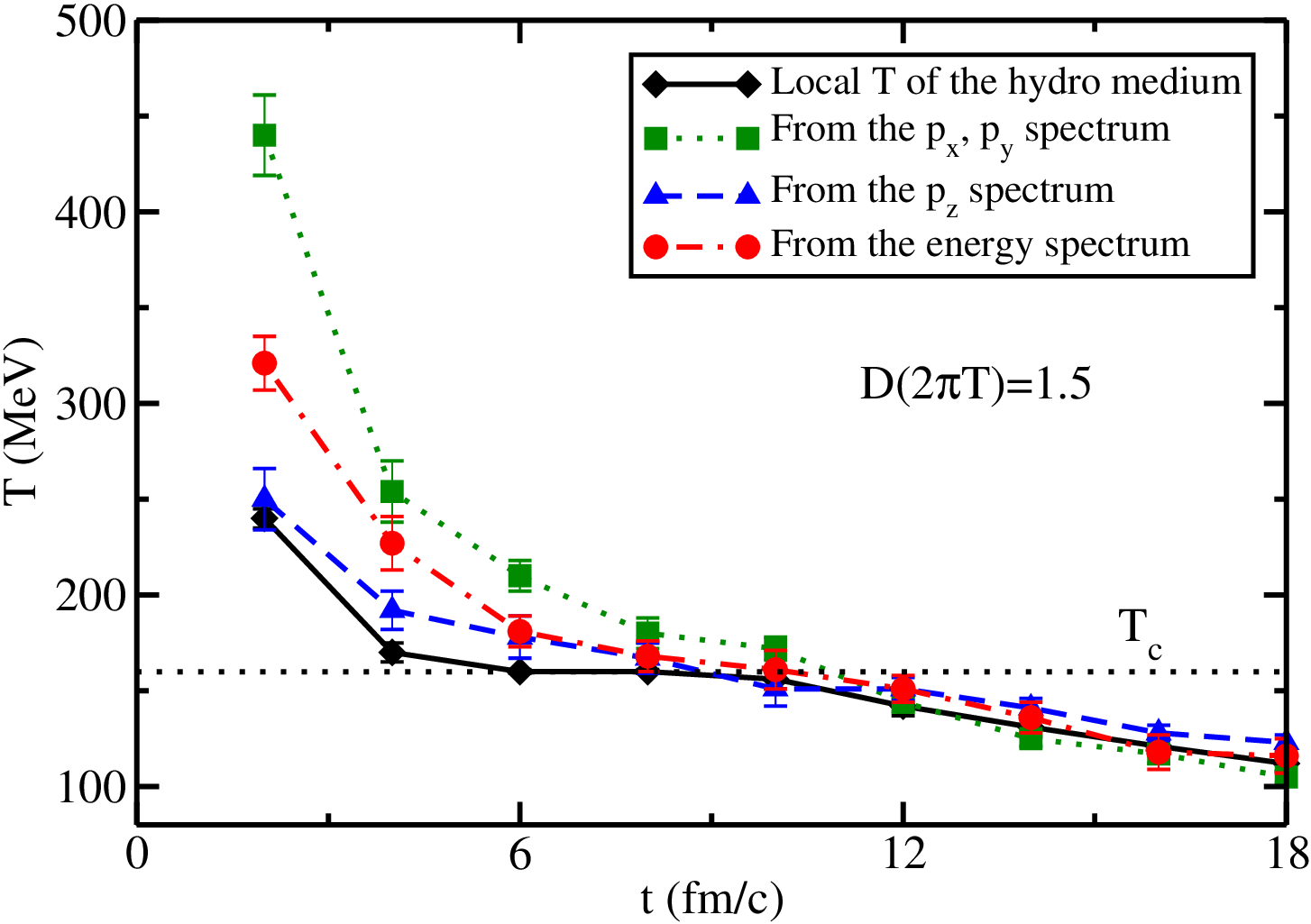}
\label{cascade1-5}
\end{minipage}
\caption{\label{cascade}(Color online) Thermalization process of charm quarks in a realistic QGP medium simulated via hydrodynamics calculation with the Glauber initial condition. The left shows the results with a diffusion coefficient of $D=6/(2\pi T)$; and the right shows the results with $D=1.5/(2\pi T)$.}
\end{figure}

Fig.\ref{cascade} displays the time evolution of the temperature parameters of charm quarks extracted from different methods, in comparison with the selected temperature band of the QGP medium. With a diffusion coefficient of $D=6/(2\pi T)$, the ``temperature" of the charm quarks never manages to catch up with that of the medium even until  freeze-out. A closer observation indicates that the charm quarks remain far off equilibrium during the entire lifetime of the QGP phase, i.e., when the medium temperature is above $T_c$. However, if the diffusion coefficient is reduced artificially to $D=1.5/(2\pi T)$, the thermalization process can be accelerated significantly and the ``temperature" of the charm quarks is able to catch up with that of the medium during its QGP phase. This result suggests that with the current adopted diffusion coefficient $D=6/(2\pi T)$ that describes experimental data within the framework of Langevin equation, charm quarks may remain far off equilibrium during the QGP lifetime. The strong coupling strength, as revealed by the heavy flavor suppression and flow behavior, is insufficient to conclude that charm quarks are able to thermalize with the QGP medium.

Of course, the thermalization speed of charm quarks depends on various factors such as the their energy scale. In \cite{Cao:2011et}, we compare the thermalization process of charm quarks with different initial momenta and showed that charm quarks with initial momentum below 1~GeV/c can thermalize fast enough within the QGP temperature range and during the QGP lifetime. To the contrary, it takes much longer time for them to thermalize with an initial momentum above 3~GeV/c. Similar comparisons between different medium temperatures and diffusion coefficients are also available in \cite{Cao:2011et}.

\section{Model and parameter dependence of final state heavy flavor spectra}
\label{modeldep}

In this section, we systematically explore how the final state heavy flavor spectra depend on different phenomenological models and parameters.
We start with the influence of geometric vs. flow properties of QGP medium on the heavy flavor suppression and elliptic flow. There are two major factors contributing to the collective flow $v_2$ of heavy quarks: the geometric asymmetry of the medium created in the heavy-ion collisions and the collective behavior of such medium. Because of the asymmetric geometry of the QGP medium, longer paths will be traversed by heavy quarks moving in the out-of-plane ($\hat{y}$) direction than in the in-plane ($\hat{x}$) direction, where the reaction plane is defined to be spanned by the impact parameter and the beam axis directions. Thus, in the absence of medium flow, heavy quarks, after passing through such anisotropic medium, would have larger momentum in the  $\hat{x}$ direction than in the $\hat{y}$ direction, $\langle p_x^2 \rangle > \langle p_y^2 \rangle$, resulting in a positive elliptic flow. Meanwhile, the collective motion of the medium also contributes positively to heavy quark elliptic flow since the push of the radial flow is more prominent in the $\hat{x}$ direction. Therefore, the total elliptic flow developed during the propagation of heavy quarks in such an anisotropic hydrodynamic medium is due to a combination of these two factors. We are able to decouple the effects of these two factors by solving Langevin equation in the global c.m. frame instead of the local rest frame of QGP, so that the contribution from medium flow is shielded.

\begin{figure}[htb]
\begin{minipage}{18pc}
\includegraphics[width=18pc,clip=]{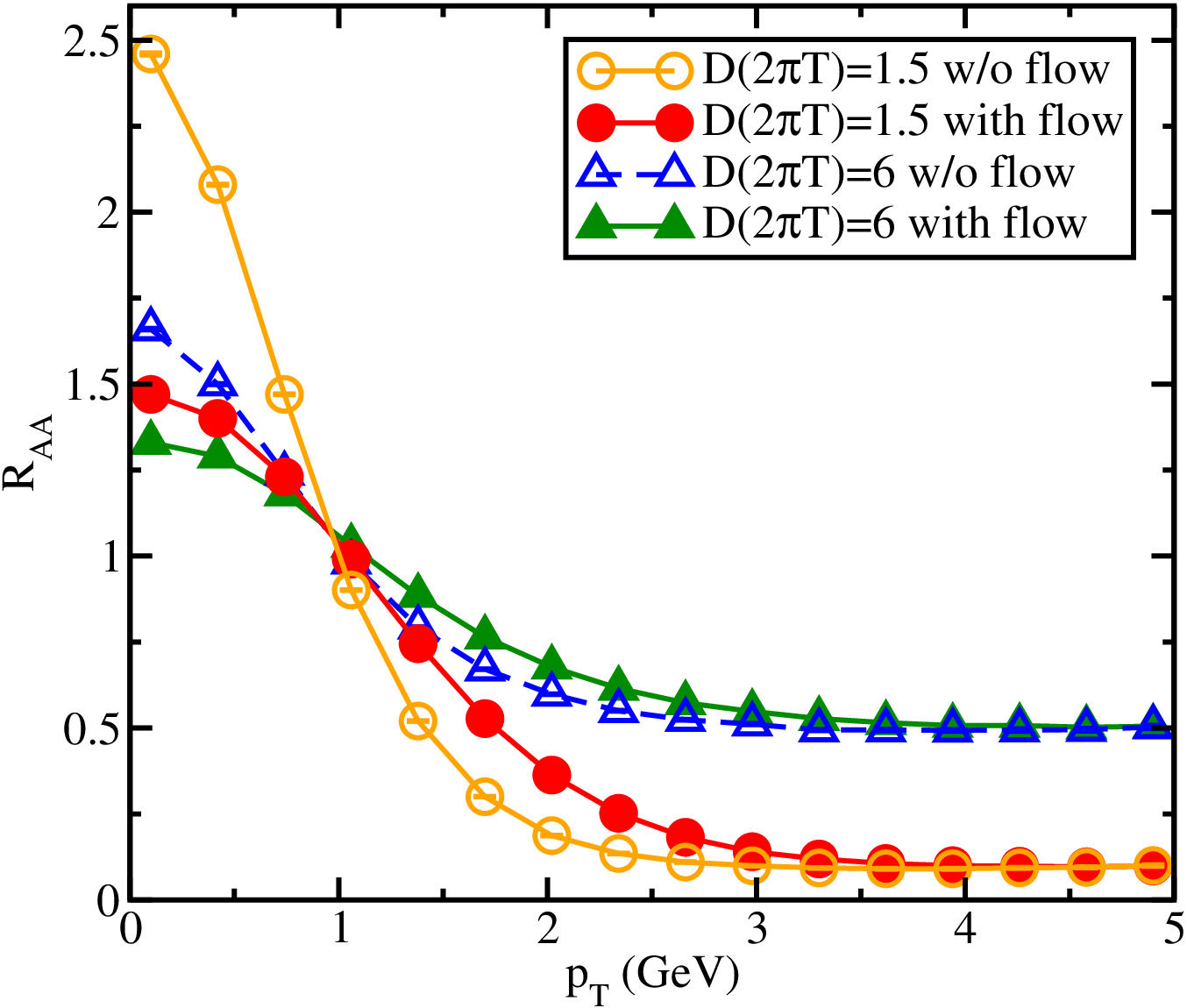}
\end{minipage}\hspace{2pc}
\begin{minipage}{14pc}
\includegraphics[width=18pc,clip=]{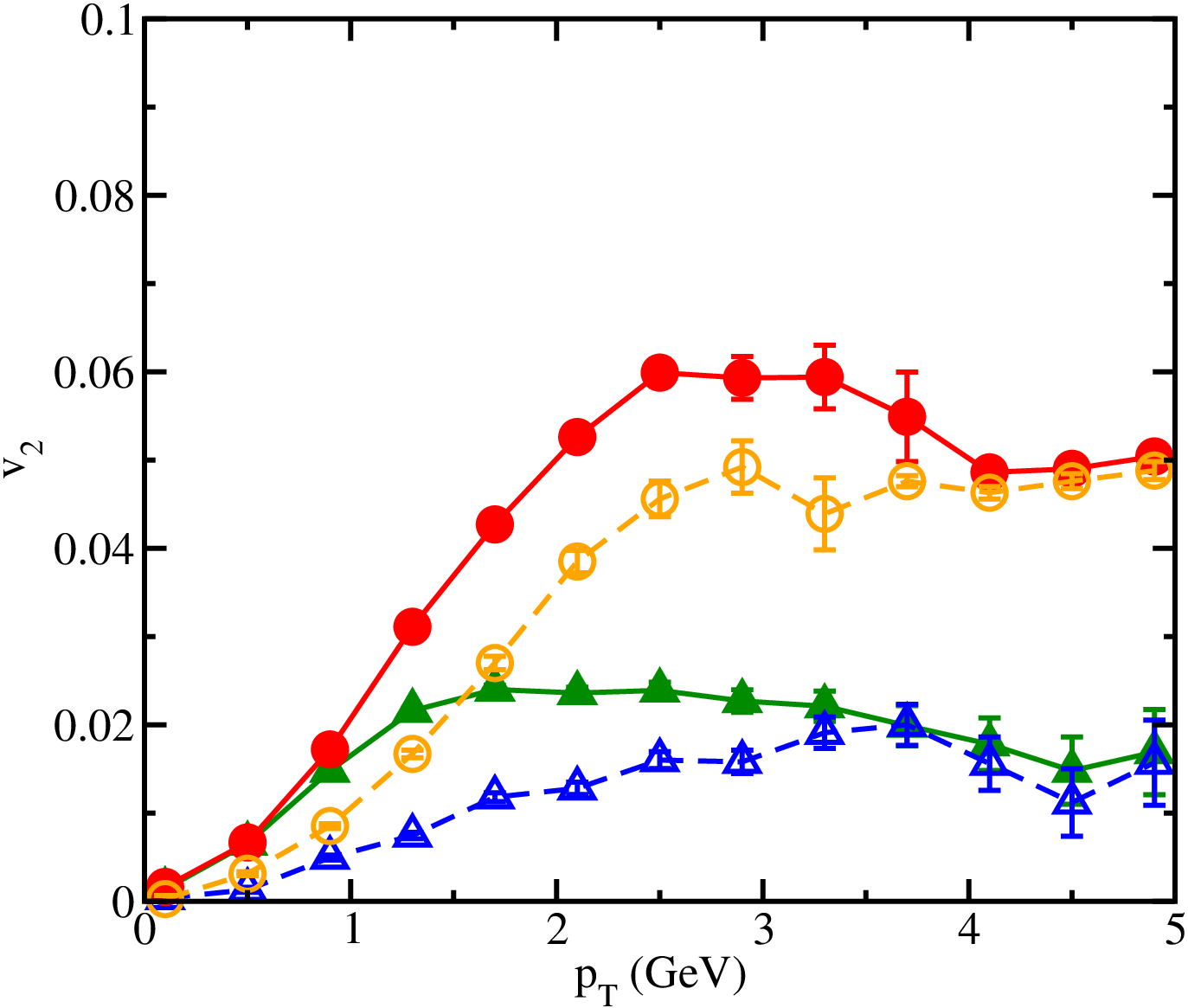}
\end{minipage}
\caption{\label{floweffect}(Color online) A comparison between the influence of QGP media with and without collective flow on $R_{AA}$ and $v_2$ of charm quarks. Both media are generated with the Glauber initial condition.}
\end{figure}

Fig.\ref{floweffect} compares the charm quark $R_{AA}$ and $v_2$ with and without this flow effect after traversing a QGP medium created by Au-Au collisions with an impact parameter of 6.5~fm. Both the $R_{AA}$ and the $v_2$ spectra shown here indicate that the medium geometry dominates the high $p_T$ region, while the collective motion of the medium has a significant impact in the low $p_T$ region. In the limit of small $p_T$, heavy quarks thermalize with the medium during the QGP lifetime, and therefore, their flow behavior would entirely follow the medium motion.

\begin{figure}[htb]
\begin{minipage}{18pc}
\includegraphics[width=18pc,clip=]{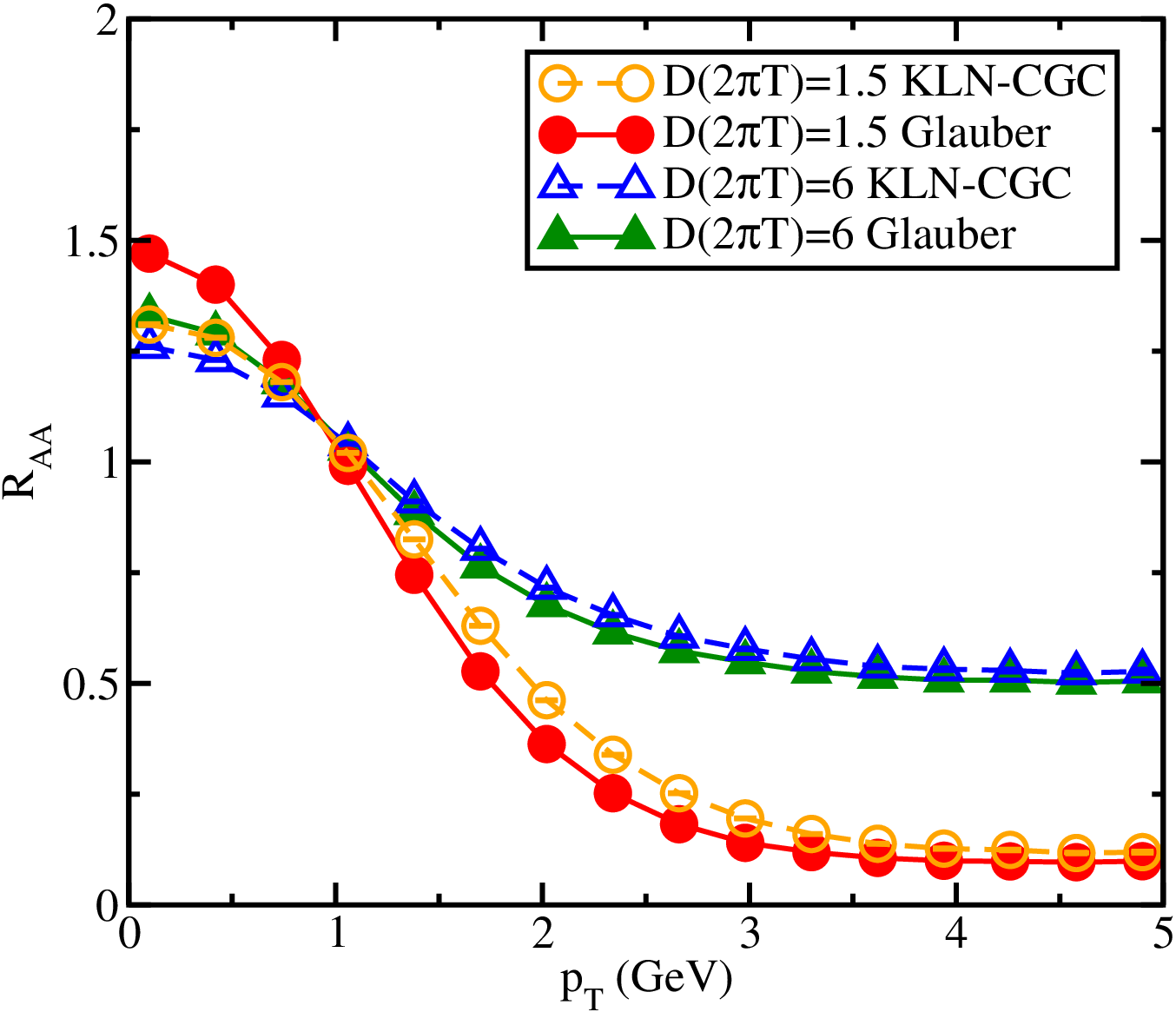}
\end{minipage}\hspace{2pc}
\begin{minipage}{14pc}
\includegraphics[width=18pc,clip=]{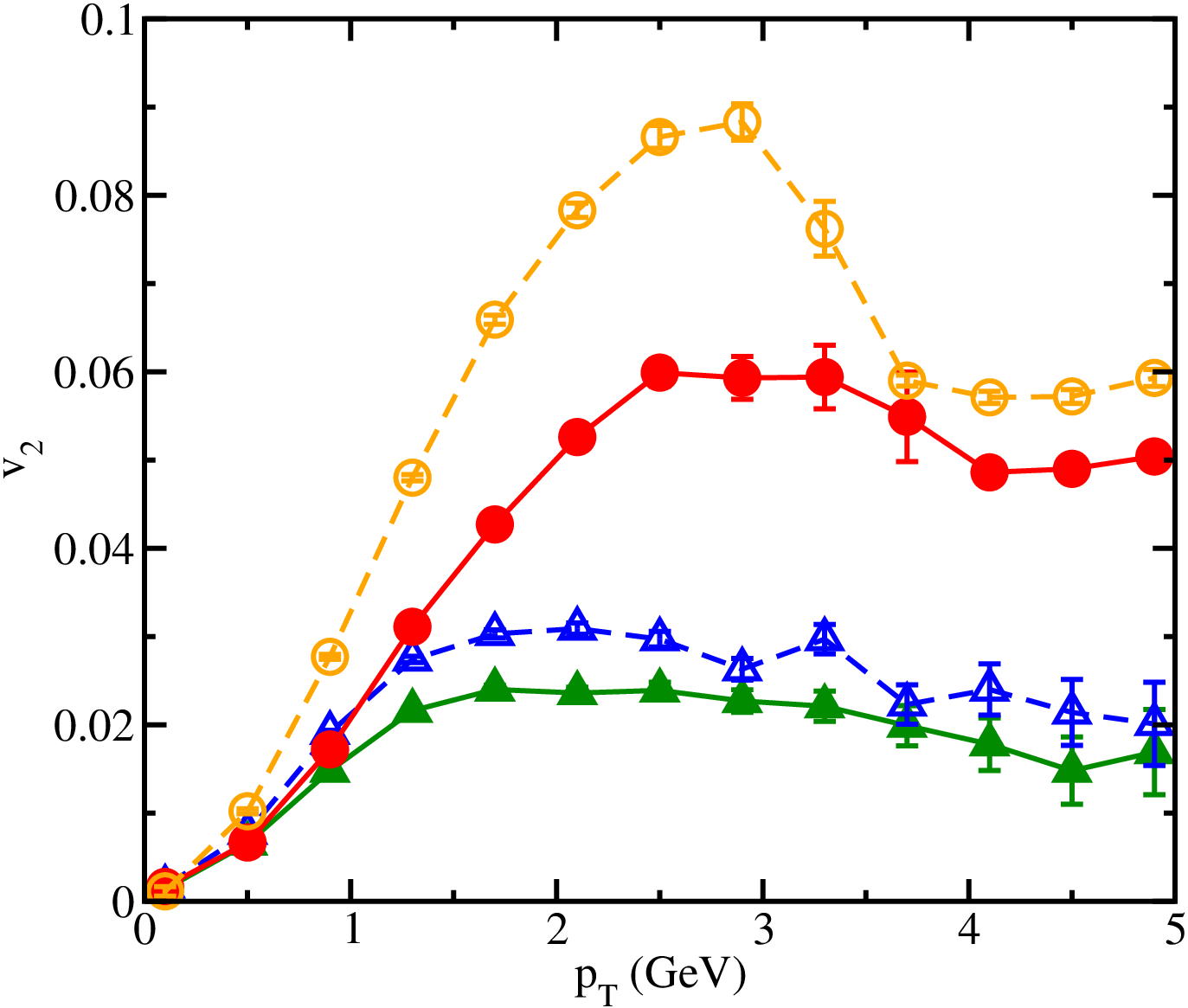}
\end{minipage}
\caption{\label{hydroeffect}(Color online) A comparison between the influence of QGP media with the Glauber and the KLN-CGC initial conditions on $R_{AA}$ and $v_2$ of charm quarks.}
\end{figure}

We may further investigate the effect of the spatial medium distribution  on the heavy quark energy loss  and the development of heavy quark  elliptic flow by utilizing different initial conditions for the hydrodynamic simulation of the QGP. Fig.\ref{hydroeffect} compares two initial conditions of the hydrodynamic evolution. Since the KLN-CGC initial condition exhibits a larger medium eccentricity ($\epsilon_2 = \langle y^2-x^2 \rangle / \langle y^2 + x^2 \rangle$)  than the Glauber initial condition, and as discussed above, such geometric factors dominate the high $p_T$ region, the KLN-CGC initial condition leads to a significantly higher $v_2$ of charm quarks than Glauber from medium to high $p_T$ region. On the other hand, the overall suppression of heavy quark is not apparently influenced by the choice of initial conditions.

Last but not least, we examine charm vs. bottom contribution to the spectrum of heavy flavor decay electron. The uncertainty in relative normalization of charm and bottom quark production from current pQCD calculation will lead to an uncertainty in the spectrum of the decayed electrons. Here, we treat the ratio of charm and bottom quarks as a free parameter for our calculation, and investigate how
the variation of this ratio affects the final non-photonic electron distributions.

\begin{figure}[htb]
\begin{minipage}{18pc}
\includegraphics[width=18pc,clip=]{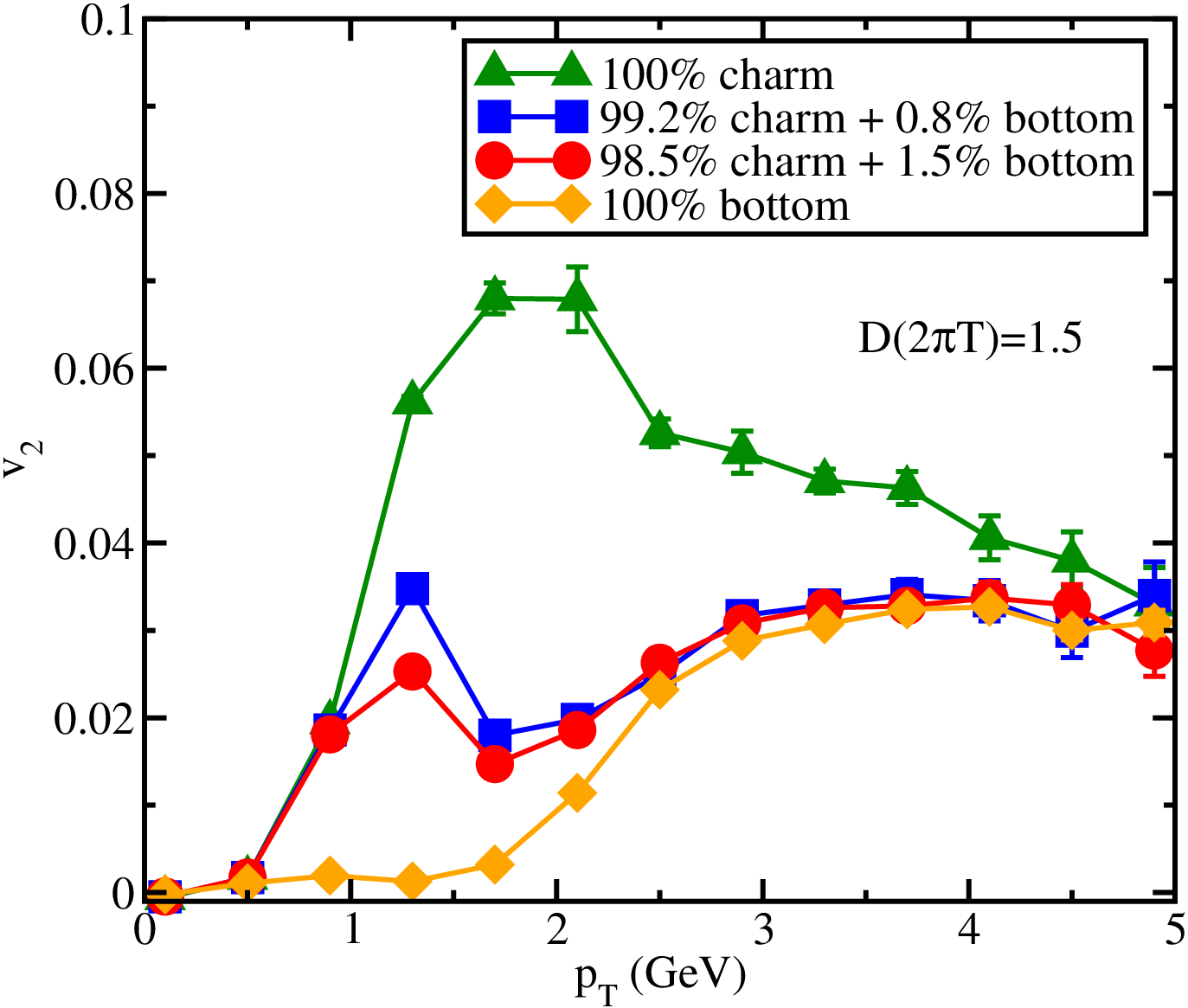}
\end{minipage}\hspace{2pc}
\begin{minipage}{14pc}
\includegraphics[width=18pc,clip=]{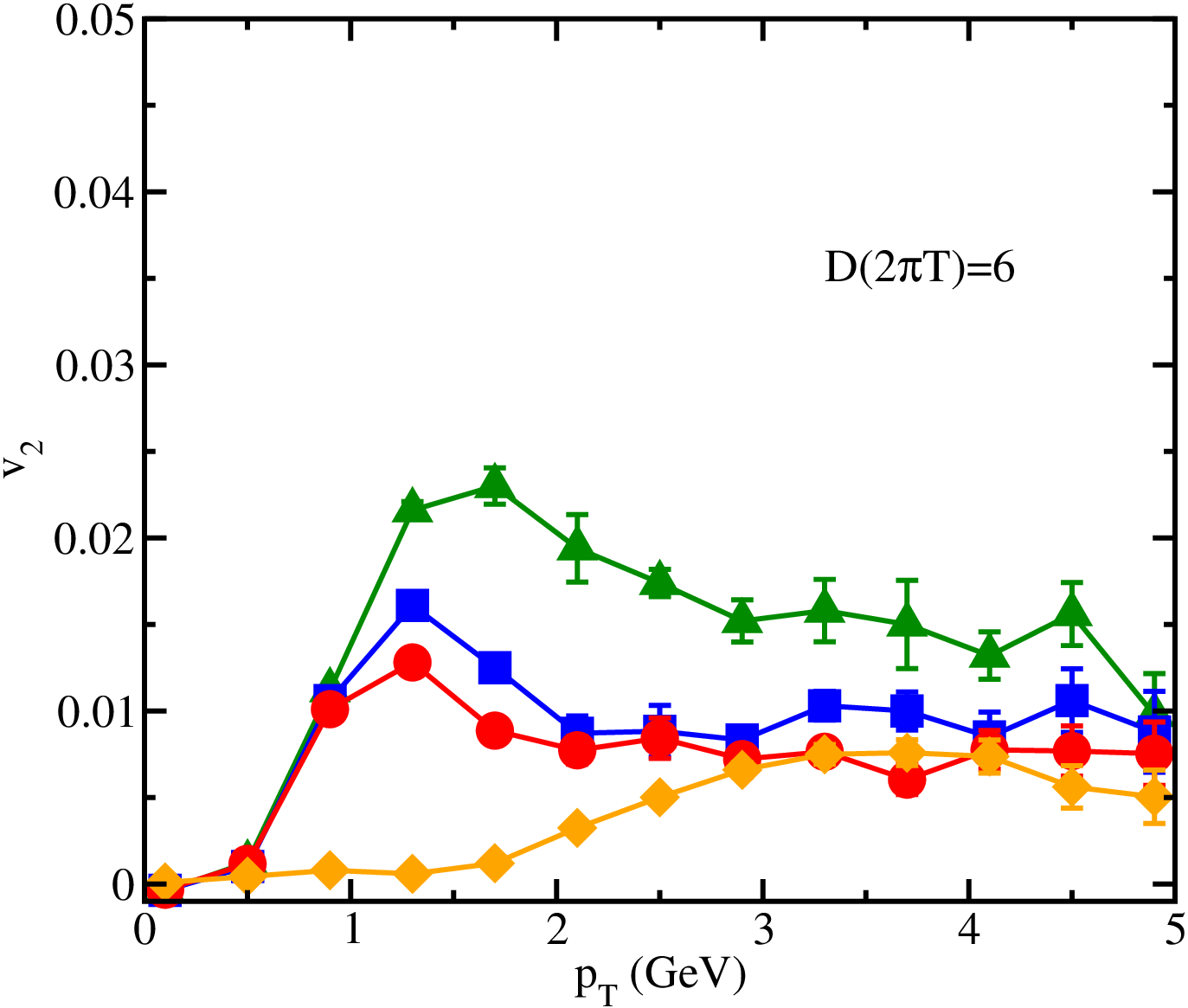}
\end{minipage}
\caption{\label{cbratio}(Color online) A comparison of non-photonic electron $v_2$ between different initial charm/bottom ratios. The left figure is for a diffusion coefficient of $D=1.5/(2\pi T)$, and the right for $D=6/(2\pi T)$. And the KLN-CGC initialization is adopted for the hydrodynamics calculation here.}
\end{figure}

Four initial normalizations of charm vs. bottom quarks are chosen and compared in Fig.\ref{cbratio} -- pure charm, pure bottom, and two mixtures: 99.2\% charm quarks with 0.8\% bottom quarks, and 98.5\% charm quarks with 1.5\% bottom quarks. As shown in \cite{Armesto:2005mz}, the bottom quark contribution to the
electron spectra may start dominating over the charm quark contribution at transverse momenta as low as 3~GeV or as high as 9~GeV. Our two mixtures of charm and bottom quarks have about a factor of 2 difference in their ratio, representing an estimate of the uncertainties due to our limited control on the proton-proton baseline.

One observes from Fig.\ref{cbratio} that the elliptic flow $v_2$ of heavy flavor decay electrons are very different
between the pure charm and pure bottom quark scenario. Bottom quarks lose less energy in QGP than charm quarks due to the mass hierarchy and therefore display smaller $v_2$ in the end. The results for both diffusion coefficients indicate the electron suppression factor follows charm quark behavior at low $p_T$ region, but bottom quark at high $p_T$ region. The electron spectrum is rather sensitive to the charm vs. bottom normalization especially at the peak value of $v_2$ -- a difference of $0.7\%$ in the mixing ratio can lead to a variation of approximately 25\% in electron $v_2$ for a diffusion coefficient of $D=6/(2\pi T)$ and over 30\% for $D=1.5/(2\pi T)$. Furthermore, due to the different behavior of charm vs. bottom decay electrons, the electrons from a mixture of charm and bottom quarks exhibit a very rich structure. In the intermediate $p_T$ region where the transition from charm dominance to bottom dominance takes place, a  non-monotonic $p_T$ dependence (a peak-dip structure) of $v_2$ is observed. Such behavior is more prominent for the smaller value of the diffusion coefficient $D=1.5/(2\pi T)$ (left) and disappears for larger value $D=6/(2\pi T)$ (right), since a smaller diffusion coefficient increases the interaction strength with the medium and thus the energy loss of charm quarks and their elliptic flow, while such an enhancement is far less for bottom quarks due to their larger mass. Current experimental results seem not able to determine whether such a peak-dip structure is present or not in the non-photonic electron spectrum due to the large experimental error bars. Further improvement of measurement would be helpful for extracting the diffusion coefficient and therefore the coupling strength between heavy quarks and the QGP medium.

\section{Summary and outlook}
\label{sumandout}

We have studied the energy loss and thermalization of heavy quarks in both infinite and finite QGP matter in the framework of Langevin approach. 
We have established a rigorous criterion to test for the thermalization of heavy quarks via extracting and comparing the temperature parameters of heavy quarks from both their energy and momentum spectra. The equilibrium of heavy quarks is defined as the state in which the temperature parameters extracted from different methods agree with each other and approach that of the traversed medium.

Our results indicate that with the current adopted diffusion coefficient, heavy quarks are only partially thermalized with the QGP medium. 
The presence of a strongly interacting system (as revealed by the heavy flavor $R_{AA}$ and $v_2$) is insufficient to conclude that heavy quarks have actually thermalized in the medium. In fact, heavy quarks may remain off equilibrium during the entire QGP lifetime.

In addition, we have investigated how the final state spectra are affected by various components of the phenomenological studies, such as the geometry and the collective flow of the hydrodynamic medium, the initial production ratio of charm to bottom quarks, and the coupling strength between heavy quarks and medium.
It is found that the geometric anisotropy of the medium dominates the final heavy quark evolution and flow in the high $p_T$ region, while the collective flow of the medium dominates the low $p_T$ region. The impact of the QGP geometry on the heavy quark energy loss has also been explored by comparing the Glauber and the KLN-CGC initializations for the hydrodynamic medium -- while the overall suppression is not apparently influenced by the choice of initial conditions, the KLN-CGC model provides a significantly higher heavy flavor $v_2$ because of its larger eccentricity for the QGP profile than the Glauber.  Furthermore, the sensitivity of the final state electron spectra to the relative charm vs. bottom normalization has also been studied -- a less than $1\%$ difference in the initial charm-to-bottom ratio can lead to more than $30\%$ variation of the non-photonic electron $v_2$. Therefore, narrowing down these uncertainties is useful for a better understanding of the interaction dynamics between heavy quarks and the QGP medium.

Our study constitutes an essential step towards more quantitatively understanding the evolution and modification of heavy quark properties in a hot and dense QCD medium. 
The current application can be further improved in several directions.
For example, both the turbulent chromo-electromagnetic fields in the pre-equilibrium stage before QGP \cite{Mrowczynski:1993qm} and the interactions of heavy flavor mesons with the hadronic matter after hadronization \cite{He:2012xz} may have non-negligible contributions to the final state observables.
Another interesting effect on the heavy quark evolution in dense nuclear medium is the radiative energy loss induced by multiple scatterings, which may be important to include, especially moving from the current RHIC energy level to the LHC era. Such effect has been discussed in other frameworks \cite{Gossiaux:2010yx}, but is still absent in the current Langevin algorithm. These questions will be addressed in our future studies. 

\section*{References}

\bibliography{SCrefs}

\end{document}